\documentclass[preprint,pre,onecolumn]{revtex4}
\usepackage{amsfonts}
\usepackage{amsmath}
\usepackage{amssymb}
\usepackage{graphicx}

\begin{document}

\preprint{}
\title{Colloids with key-lock interactions: non-exponential relaxation,
aging and anomalous diffusion}
\author{Nicholas A. Licata and Alexei V. Tkachenko}
\affiliation{Department of Physics and Michigan Center for Theoretical Physics,
University of Michigan, 450 Church Street, Ann Arbor, Michigan 48109}

\begin{abstract}
The dynamics of particles interacting by key-lock binding of attached
biomolecules are studied theoretically. \ Experimental realizations of such
systems include colloids grafted with complementary single-stranded DNA
(ssDNA), and particles grafted with antibodies to cell-membrane proteins. \
Depending on the coverage of the functional groups, we predict two distinct
regimes. \ In the low coverage \textit{localized regime}, there is an
exponential distribution of departure times. \ As the coverage is increased
the system enters a \textit{diffusive regime} resulting from the interplay
of particle desorption and diffusion. \ This interplay leads to much longer
bound state lifetimes, a phenomenon qualitatively similar to \textit{aging}
in glassy systems. \ The diffusion behavior is analogous to dispersive
transport in disordered semiconductors: depending on the interaction
parameters it may range from a finite renormalization of the diffusion
coefficient to anomalous, \textit{subdiffusive} behavior. \ We make
connections to recent experiments and discuss the implications for future
studies. \ 
\end{abstract}

\maketitle

\section{Introduction}

In this paper we present a theoretical study of desorption and diffusion of
particles which interact through key-lock binding of attached biomolecules.
\ It is becoming common practice to functionalize colloidal particles with
single-stranded DNA (ssDNA) to achieve specific, controllable interactions (%
\cite{crocker},\cite{chaikin},\cite{micelle},\cite{natreview},\cite{rational}%
,\cite{synthesis}). \ Beyond the conceptual interest as a model system to
study glassiness\cite{licata} and crystallization, there are a number of
practical applications. \ Colloidal self-assembly may provide a fabrication
technique for photonic band gap materials (\cite{photonic},\cite{wiremesh}).
\ One of the major experimental goals in this line of research is the
self-assembly of colloidal crystals using DNA mediated interactions. \ The
difficulty stems in part from the slow relaxation dynamics in these systems.
\ The main goal of this paper is to understand how the collective character
of key-lock binding influences the particle dynamics. \ In doing so we gain
valuable insight into the relaxation dynamics, and propose a modified
experimental setup whose fast relaxation should facilitate colloidal
crystallization. \ 

Similar systems have also attracted substantial attention in other areas of
nanoscience. \ In particular, by functionalizing nanoparticles with
antibodies to a particular protein, the nanoparticles have potential
applications as smart, cell-specific drug delivery vehicles (\cite%
{membranebend},\cite{dendrimer}). \ These nanodevices take advantage of the
fact that certain cancerous cells overexpress cell membrane proteins, for
example the folate receptor. \ An improved understanding of desorption and
diffusion on the cell membrane surface may have implications for optimizing
the design of these drug delivery vehicles. \ 

In what follows we present our results on the dynamics of particles which
interact through reversible key-lock binding. \ The plan for the paper is
the following. \ In section II we introduce the key-lock model and explain
the origin of the two model parameters $\Delta $ and $\overline{m}$. \ The
parameter $\Delta $ determines the binding energy for the formation of a
key-lock pair. \ The parameter $\overline{m}$ is the mean of the
distribution for the number of key-lock bridges. \ Depending on $\overline{m}
$, which is related to the coverage of the functional groups (e.g. ssDNA),
there are two distinct regimes. \ At low coverage there is an exponential
distribution of departure times, but no true lateral diffusion. \ As the
coverage increases, we enter a regime where the particle dynamics is a
result of the interplay between desorption and diffusion. \ An estimate is
provided for the value of $\overline{m}$ which determines the crossover from
the localized to diffusive regime. \ In section III the localized regime is
discussed in detail. \ In this regime the particle is attached to a finite
cluster and remains localized near its original location until departing. \
We derive the partition function for the finite clusters, and calculate the
departure time distribution. \ In section IV we determine the departure time
distribution in the diffusive regime. \ We present an effective Arrhenius
approximation for the hopping process and a Fourier transform method which
greatly simplifies the calculation. \ In section V we discuss the random
walk statistics for the particles' in-plane diffusion. \ A set of parametric
equations is derived to relate the average diffusion time to the mean
squared displacement. \ The lateral motion is analogous to dispersive
transport in disordered semiconductors, ranging from standard diffusion with
a renormalized diffusion coefficient to anomalous, subdiffusive behavior. \
In section VI we connect our results to recent experiments with DNA-grafted
colloids. \ We then discuss the implications of the work for designing an
experiment which facilitates faster colloidal crystallization. \ In section
VII we conclude by summarizing our main results. \ 

\begin{figure}[tbp]
\includegraphics[width=4.6112in,height=3.4705in]{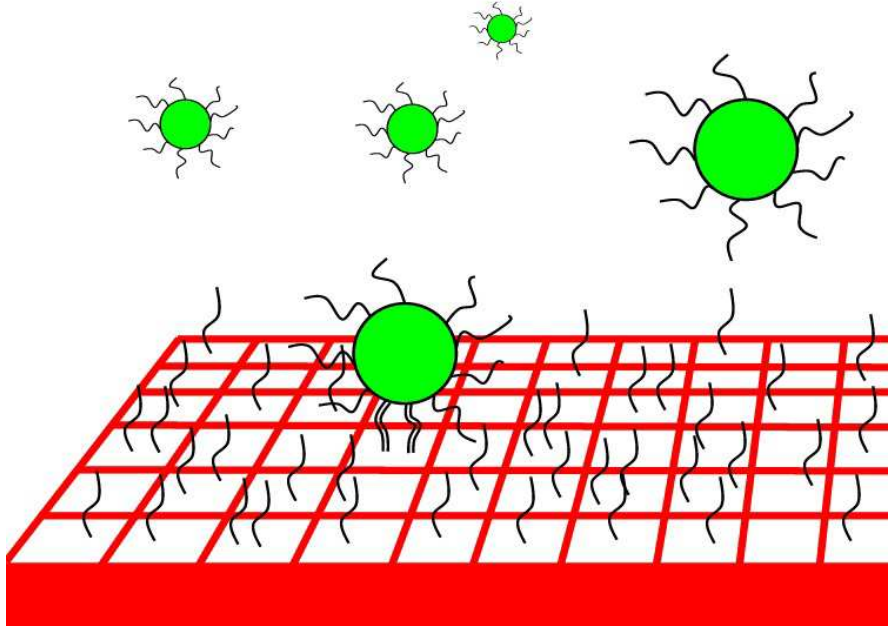}
\caption{(Color online). Graphical depiction of particles interacting with a
flat, 2D substrate by multiple key-lock binding. }
\label{substrate}
\end{figure}

\section{Model}

We now present the model, where a single particle interacts with a flat
two-dimensional surface by multiple key lock binding (see Fig. \ref%
{substrate}). \ At each location on the surface there are $m$ key-lock
bridges which may be open or closed, with a binding energy of $\epsilon $
for each key-lock pair. \ Here we have neglected the variation in $\epsilon $%
. \ In the case of the DNA-colloidal system mentioned in the introduction,
the model parameter $\epsilon $ is related to the hybridization free energy
of the DNA. \ The resulting $m$-bridge free energy plays the role of an
effective local potential for the particle\cite{statmech}:%
\begin{align}
U(m)& =-k_{B}Tm\Delta  \label{potenergy} \\
\Delta & \equiv \log (1+\exp [\epsilon /k_{B}T])
\end{align}%
Generically, $m$ is a Poisson distributed random number $P_{m}=\overline{m}%
^{m}\exp (-\overline{m})/m!$ where $\overline{m}$ denotes the mean of the
distribution. \ The model parameter $\overline{m}$ is a collective property
of the particle-surface system. \ For example, consider the case of
dendrimers functionalized with folic acid, which can be utilized for
targeted, cell specific chemotherapy. \ The folic acid on the dendrimer
branch ends form key-lock bridges with folate receptors in the
cell-membrane. \ In this case $\overline{m}$ will depend on the distribution
of keys (folic acids) on the dendrimer, and the surface coverage of locks
(folate receptors) in the cell membrane. \ 

At each location, the particle is attached to the surface by $m$ bridges. \
To detach from the surface the particle must break all its connections, in
which case it departs and diffuses away into solution. \ Alternatively the
particle can hop a distance $a$ to a new location characterized by a new
value of the bridge number $m$. \ By introducing the correlation length $a$,
we have coarse-grained the particle motion by the distance after which the
new value of the bridge number becomes statistically independent of the
value at the previous location. \ In the localized regime the particle
remains close to its original location until departing. \ In the diffusive
regime the particle is able to fully explore the surface through a random
walk by multiple breaking and reforming of bridges. \ 

\begin{figure}[tbp]
\includegraphics[width=4.6138in,height=3.4714in]{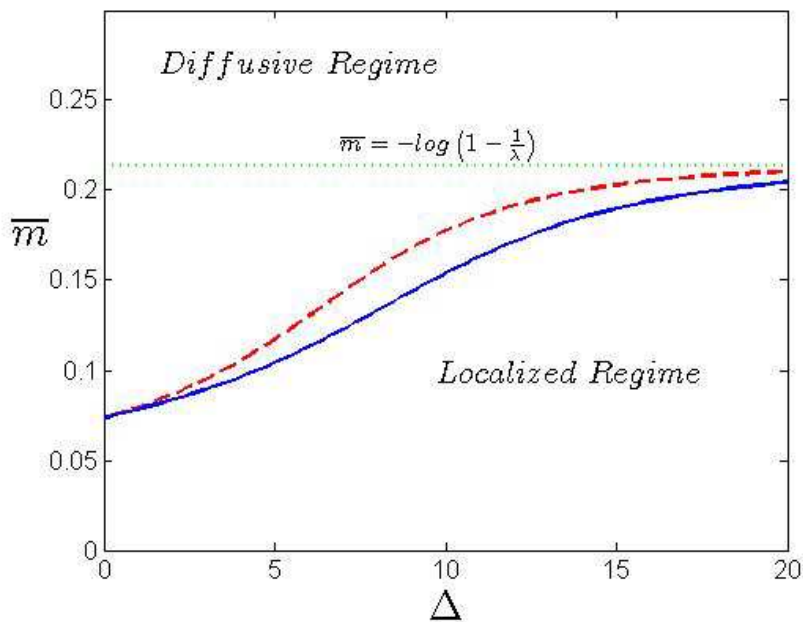}
\caption{(Color online).  \ The crossover
from the localized to the diffusive regime below the percolation threshold.
\ Estimates based on the characteristic cluster size (Eq. \protect\ref{crossnum}, dashed line) and 
confinement of the random walk (Eq. \protect\ref{crosswalk}, solid line)
give similar crossovers. \ For large $\Delta $ the crossover condition is $\overline{m}=
-\log \left( 1-\frac{1}{\protect\lambda }\right) $. \ }
\label{crosspoint3}
\end{figure}

Naively one might expect the
crossover between the two regimes to occur at the percolation threshold,
where one first encounters an infinitely connected cluster of sites with $m>0
$. \ However, the crossover from the localized to diffusive regime occurs at
smaller $\overline{m}$\ than predicted by percolation theory. \ If $p_{c}=1/2
$ denotes the critical probability for site percolation on the triangular
lattice, the percolation transition occurs at $\overline{m}=\log \left(
2\right) $. \ There are two alternative estimates for the crossover from the
localized to the diffusive regime. \ The first is to compare the average
number of steps $n=\exp \left( \Delta \overline{m}\right) $ the particle
takes before departing (see section IV) to the characteristic cluster size $%
s_{c}=1/\log (1/\lambda p)$ below the percolation threshold. \ Here $\lambda
=5.19$ is a numerical constant for the triangular lattice\cite{cluster}, and
in the percolation language $p=1-\exp (-\overline{m})$ is the occupancy
probability. \ The crossover condition $n=s_{c}$ can be expressed as a
function of $\overline{m}$. \ 
\begin{equation}
\Delta =-\frac{1}{\overline{m}}\log \left[ -\log \left\{ \lambda \left(
1-e^{-\overline{m}}\right) \right\} \right]   \label{crossnum}
\end{equation}%
Alternatively, in the localized regime the particles' random walk is
confined by the characteristic cluster size. \ Below percolation the radius
of gyration of the cluster is $R_{s}\sim s^{\rho }$ with $\rho =0.641$ in
two dimensions. \ Comparing the radius of gyration of the cluster to the
radius of gyration for the particles' random walk, the crossover occurs at $%
n=\left( s_{c}\right) ^{2\rho }$. \ 
\begin{equation}
\Delta =-\frac{2\rho }{\overline{m}}\log \left[ -\log \left\{ \lambda \left(
1-e^{-\overline{m}}\right) \right\} \right]   \label{crosswalk}
\end{equation}%
Since $2\rho $ differs from $1$ by less than $30\%$, both conditions give
similar crossovers (see Fig. \ref{crosspoint3}). \ The saturation at $%
\overline{m}=-\log \left( 1-\frac{1}{\lambda }\right) $ occurs for very
large $\Delta $, as a result for binding energies of a few $k_{B}T$ per
bridge the crossover occurs at $\overline{m}\simeq 0.1$. \ 

\section{Localized Regime}

In the percolation language, when the occupancy probability $p=1-\exp (-%
\overline{m})$ is small, particles are localized on finite clusters. \ In
this localized regime particles are able to fully explore the cluster to
which they are attached before departing. \ This thermalization of particles
with finite clusters permits an equilibrium calculation of the cluster free
energy $F=-k_{B}T\log \langle Z\rangle $. \ The departure rate is given by
the Arrhenius relation $K=\frac{1}{\tau _{0}}\exp \left( F/k_{B}T\right) $.
\ Here $\tau _{0}$ is a characteristic timescale for bridge formation. \ The
probability that the particle departs between $t$ and $t+dt$ is determined
from the departure time distribution $\Phi (t)dt\simeq K\exp [-Kt]dt$. \ 

To begin we calculate the partition function for the finite clusters. \ The
cluster is defined as $s$ connected sites on the lattice, all of which are
characterized by $0<m<m^{\ast }$ bridges. \ For Poisson distributed bridge
numbers the partition function for the finite cluster is: \ 
\begin{equation}
Z(m^{\ast },s)=\sum_{i=1}^{s}\sum_{m_{i}=1}^{m^{\ast }-1}\widetilde{P}_{m_{i}}\exp 
(\Delta m_{i})=\frac{s}{\exp (\overline{m})-1}(\exp (\overline{m}
e^{\Delta })Q(\overline{m}e^{\Delta },m^{\ast })-1)
\end{equation}
\ Because by definition the cluster does not contain sites with $m=0$
bridges we have renormalized the probability distribution $\widetilde{P}
_{m}=P_{m}/(1-\exp (-\overline{m}))$ so that $\sum_{m=1}^{\infty }\widetilde{
P}_{m}=1$. \ Here $Q(x,m^{\ast })\equiv \Gamma (x,m^{\ast })/\Gamma (m^{\ast
})=\exp (-x)\sum_{k=0}^{m^{\ast }-1}x^{k}/k!$ is the regularized upper
incomplete $\Gamma $ function. \ In the language of the statistics of
extreme events, $m^{\ast }-1$ is the maximum "expected" value of $m$ in a
sample of $s$ independent realizations\cite{dispersive}. The point is that
on finite clusters we should not expect to achieve arbitrarily large values
of the bridge number. \ Hence when averaging the partition function to
obtain the cluster free energy one should only average over sites with $
m<m^{\ast }$. \ The distribution function for $m^{\ast }$ is obtained by
noting that the probability that all $s$ values of $m$ are less than $
m^{\ast }$ is $\left( \sum_{m=1}^{m^{\ast }-1}\widetilde{P}_{m}\right)
^{s}=\left( \frac{\exp (\overline{m})Q(\overline{m},m^{\ast })-1}{\exp (
\overline{m})-1}\right) ^{s}$. \ By differentiating this quantity with
respect to $m^{\ast }$ we obtain the distribution function for the maximum
expected value of $m$. \ 
\begin{equation}
f_{s}(m^{\ast })=s\left( \frac{\exp (\overline{m})Q(\overline{m},m^{\ast })-1
}{\exp (\overline{m})-1}\right) ^{s-1}\widetilde{P}_{m^{\ast }}
\end{equation}
\ The cluster size distribution below the percolation threshold is
exponential\cite{percolationtheory} with characteristic cluster size $
s_{c}=1/\log (1/\lambda p)$. \ 
\begin{equation}
p_{s}(p)=\frac{1-\lambda p}{\lambda }\exp \left( -\frac{s}{s_{c}}\right)
\end{equation}
The summation over $s$ can be performed analytically, which allows the
result to be expressed as a single summation over $m^{\ast }$. \ 
\begin{equation}
\langle Z\rangle =\sum_{m^{\ast }=2}^{\infty }\sum_{s=1}^{\infty
}p_{s}(p)f_{s}(m^{\ast })Z(m^{\ast },s)=
\end{equation}
\begin{equation}
\frac{\lambda (1-\lambda (1-\exp (-\overline{m})))}{\exp (\overline{m})-1}
\sum_{m^{\ast }=2}^{\infty }\widetilde{P}_{m^{\ast }}(\exp (\overline{m}
e^{\Delta })Q(\overline{m}e^{\Delta },m^{\ast })-1)\frac{1+y(m^{\ast })}{
(1-y(m^{\ast }))^{3}}  \notag
\end{equation}
\begin{equation}
y(m^{\ast })=\lambda (Q(\overline{m},m^{\ast })-\exp (-\overline{m}))
\end{equation}

\begin{figure}[tbp]
\includegraphics[width=4.6138in,height=3.4714in]{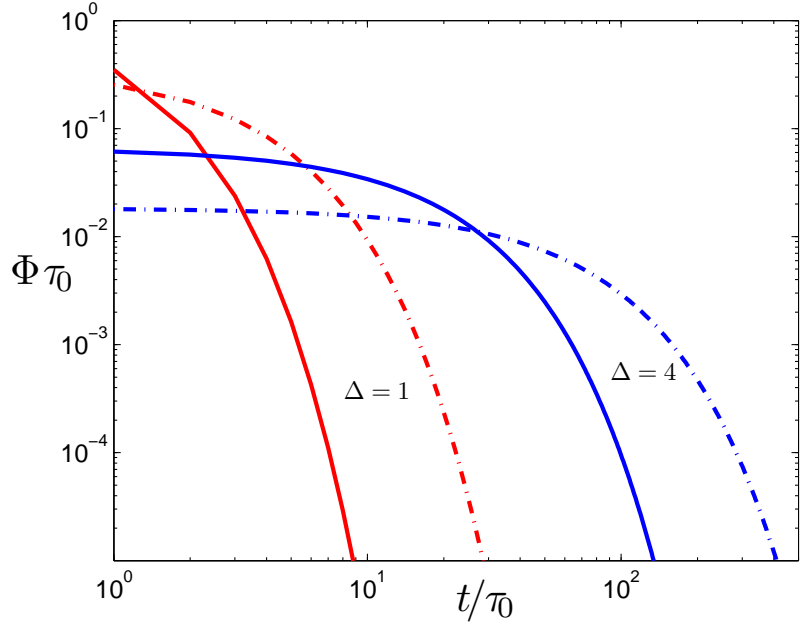}
\caption{(Color online).  \ Departure time
distribution function versus time in the localized regime with $\overline{m}
=0.05$.  The results of our calculation (solid lines) are compared to single exponential relaxation
with departure rate $K=\frac{\exp(-\Delta)}{\tau_{0}}$ (dotted lines).  }
\label{localreg3}
\end{figure}

For fixed $\overline{m}$, as in the plot (see Fig. \ref{localreg3}), changing $
\Delta $ is directly related to a change in the average binding free energy.
\ Increasing $\Delta $ leads to a reduction in the rate of particle
departure. \ 

\section{Diffusive Regime}

The departure time distribution changes significantly in the diffusive
regime. \ In this regime the particle can explore the surface to find a more
favorable connection site, which leads to a longer lifetime for the bound
state. \ This phenomenon is qualitatively similar to \textit{aging} in
glassy systems. \ In these systems one finds that the response to an
external field is time dependent\cite{glassybook}. \ In the magnetic analogy
this leads to a time dependence of the magnetization. \ Below the glass
temperature, the longer one waits before applying the external magnetic
field, the more time the system has to settle into deep energy wells, and
the smaller the response. \ In our case, the diffusive exploration of the
particle allows it to find a deeper energy well, which leads to an increase
in the bound state lifetime. \ 

\begin{figure}[tbp]
\includegraphics[width=4.6138in,height=3.4714in]{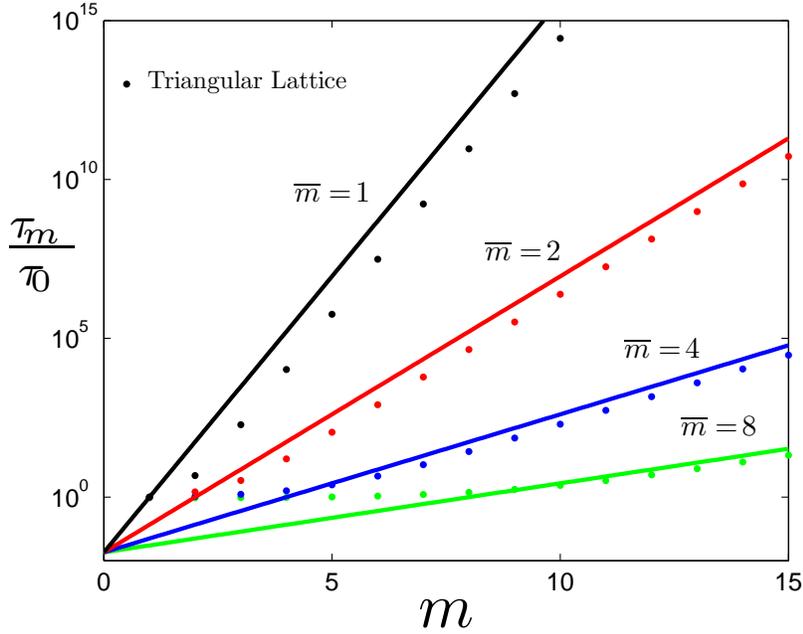}
\caption{(Color online).  Comparison of
the ensemble averaged dwell time in a lattice model (Eq. \protect\ref
{hoplattice}) to the effective Arrhenius approximation (Eq. \protect\ref{arrhenius}). 
\ In the plot the product $\Delta \overline{m}=4$ is held
constant. }
\label{tauplot6}
\end{figure}

As a result, the departure time distribution must now reflect not only
desorption, but also hopping to adjacent sites. \ The hopping rate between
neighboring sites $i$ and $j$ is given by an Arrhenius law $\kappa
_{i\rightarrow j}=\frac{1}{\tau _{0}}\exp [-\Delta (m_{i}-m_{j})\theta
(m_{i}-m_{j})]$, with $\theta (x)$ the Heaviside step function. \ In a
lattice model with coordination number $z$ the dwell time $\tau _{m}$ at a
site with $m$ bridges is calculated by averaging over the hopping rates to
the nearest neighbors (see Fig. \ref{tauplot6}). \ 

\begin{equation}
\tau_{m}=\left\langle \frac{1}{\frac{1}{z}\sum_{i=1}^{z}\kappa_{m\rightarrow
i}}\right\rangle
_{m_{1}...m_{z}}=z\sum_{m_{1}=1}^{\infty}\cdots\sum_{m_{z}=1}^{\infty}%
\widetilde{P}_{m_{1}}\cdots\widetilde{P}_{m_{z}}\frac{1}{\sum_{i=1}^{z}%
\kappa_{m\rightarrow i}}  \label{hoplattice}
\end{equation}
Fortunately, this ensemble averaging procedure can be accurately
approximated by an effective Arrhenius relation:%
\begin{equation}
\tau_{m}=\tau_{0}\exp[\Delta(m-\overline{m})]  \label{arrhenius}
\end{equation}
The validity of the approximation is most important for sites with $m\geq%
\overline{m}$ bridges, since the diffusive exploration allows the particle
to quickly cascade into these deep energy wells. \ 

The effective Arrhenius relation greatly simplifies the calculation, since
so long as $\Delta \overline{m}$ is sufficiently large, the probability of
the particle still being attached to the surface after an $n$ step random
walk is $\left( 1-K_{m}\tau _{m}\right) ^{n-1}=[1-\exp (-\Delta \overline{m}%
)]^{n-1}$. \ Here $K_{m}=\frac{1}{\tau _{0}}\exp (-\Delta m)$ is the
departure rate from a site with $m$ bridges. \ Interestingly, in this
approximation scheme the attachment probability is independent of the
particular bridge numbers $\{m_{1},...,m_{n}\}$ realized during the walk. \
Thus, the probability of departure $f_{n}$ after exactly $n$ steps is:%
\begin{align}
f_{n}& =[\exp (\gamma )-1]\exp (-\gamma n) \\
\gamma & \equiv -\log [1-\exp (-\Delta \overline{m})]
\end{align}%
The average number of steps for the random walk is $\sum\nolimits_{n=1}^{%
\infty }nf_{n}=\exp (\Delta \overline{m})$. \ To calculate the departure
time distribution $\Phi (t)$ we use $f_{n}$ to average over the departure
time distribution for walks with a given $n$, $\phi _{n}(t)$. \ \ 
\begin{equation}
\Phi (t)=\sum\limits_{n=1}^{\infty }f_{n}\phi _{n}(t)
\end{equation}%
\ 
\begin{equation}
\phi _{n}(t)=\prod\limits_{j=1}^{n}\left( \sum_{m_{j}=1}^{\infty }\widetilde{%
P}_{m_{j}}\int_{0}^{\infty }dt_{j}\left( \frac{-dS_{m_{j}}(t_{j})}{dt_{j}}%
\right) \right) \delta \left( t-\sum_{k=1}^{n}t_{k}\right)
\end{equation}

Here $S_{m}(t)$ is the survival probability at time $t$ for a site with $m$
bridges, used to determine the probability of departure between $t$ and $%
t+dt $. \ If there was only one hopping pathway with rate $\kappa$, we would
have $-\frac{dS}{dt}=\kappa\exp(-\kappa t)$. \ The generalization accounts
for the fact that the particle can hop to any of its $z$ neighbours, and the
probability of departure is not simply exponential. \ 
\begin{equation}
S_{m}(t)=\left( \sum\limits_{a=1}^{\infty}\widetilde{P}_{a}\exp\left[
-t\kappa_{m\rightarrow a}\right] \right) ^{z}
\end{equation}
It is convenient to Fourier transform $\phi_{n}(t)$ so that one can sum the
resulting geometric series for $\Phi(\omega)$. \ 
\begin{align}
\phi_{n}(\omega) & =\int_{-\infty}^{\infty}\phi_{n}(t)\exp[-i\omega t]%
dt=X(\omega)^{n}  \label{Xomega} \\
X(\omega) & \equiv\sum_{m=1}^{\infty}\widetilde{P}_{m}\sum_{m_{1}=1}^{%
\infty}\cdots\sum_{m_{z}=1}^{\infty}\widetilde{P}_{m_{1}}\cdots\widetilde {P}%
_{m_{z}}\left( \frac{\sum\limits_{i=1}^{z}\kappa_{m\rightarrow m_{i}}}{%
\sum\limits_{i=1}^{z}\kappa_{m\rightarrow m_{i}}+i\omega}\right) \\
\Phi(\omega) & =[\exp(\gamma)-1]\sum\limits_{n=1}^{\infty}\left[
\exp(-\gamma)X(\omega)\right] ^{n}=[\exp(\gamma)-1]\frac{X(\omega)}{%
\exp(\gamma)-X(\omega)}
\end{align}

To facilitate a simpler calculation, we employ a coarse-graining procedure
to dispense with the tensor indices $\{m,m_{1},...,m_{z}\}$ in the
definition of $X(\omega)$. \ In the summation there are many terms for which
the value of $\sum\limits_{i=1}^{z}\kappa_{m\rightarrow m_{i}}$ are equal,
but with different weight factors $\widetilde{P}_{m}\widetilde{P}%
_{m_{1}}\cdots\widetilde{P}_{m_{z}}$. \ To eliminate this degeneracy we
introduce a smooth function $f(\kappa)$ normalized according to $\int
f(\kappa)d\kappa=1$. \ 
\begin{equation}
X(\omega)\simeq\int f(\kappa)\frac{\kappa}{\kappa+i\omega}d\kappa
\end{equation}
\ 

The inverse Fourier transform is performed using the residue theorem to
obtain the final result. \ The contour integral is closed in the upper half
plane, with all the poles on the imaginary axis at $\omega =iz$. \ 
\begin{align}
\Phi (t)& =\frac{1}{2\pi }\int_{-\infty }^{\infty }\Phi (\omega )\exp
[i\omega t]d\omega =\frac{[\exp (\gamma )-1]}{2\pi }2\pi i\sum_{r=1}^{\infty
}res_{\omega =\omega _{r}}\left[ \frac{\exp [i\omega t]X(\omega )}{\exp
(\gamma )-X(\omega )}\right]  \label{deptimedistdif} \\
& =[\exp (\gamma )-1]i\sum_{r=1}^{\infty }res_{\omega =\omega _{r}}\left[ 
\frac{\exp [i\omega _{r}t]\left\{ X(\omega _{r})+(\omega -\omega _{r})\left( 
\frac{dX}{d\omega }\right) _{\omega =\omega _{r}}+\cdots \right\} }{\exp
(\gamma )-\left\{ X(\omega _{r})+(\omega -\omega _{r})\left( \frac{dX}{%
d\omega }\right) _{\omega =\omega _{r}}+\cdots \right\} }\right]  \notag \\
& =[\exp (\gamma )-1]i\sum_{r=1}^{\infty }\left[ \frac{-\exp [i\omega
_{r}t]X(\omega _{r})}{\left( \frac{dX}{d\omega }\right) _{\omega =\omega
_{r}}}\right]  \notag \\
& =\exp (\gamma )[\exp (\gamma )-1]\sum_{r=1}^{\infty }\frac{\exp (-z_{r}t)}{%
Y(z_{r})}  \notag \\
Y(z_{r})& \equiv \int f(\kappa )\frac{\kappa }{\left( \kappa -z_{r}\right)
^{2}}d\kappa
\end{align}%
\qquad Here $z_{r}$ labels the roots of the equation%
\begin{equation}
\exp (\gamma )-X(iz)=0
\end{equation}%
The benefit of the coarse-graining is now more transparent, as the residues
are all labeled by a single index $r$ as opposed to the tensor indices $%
\{m,m_{1},...,m_{z}\}$. \ 

\begin{figure}[tbp]
\includegraphics[width=4.6138in,height=3.4714in]{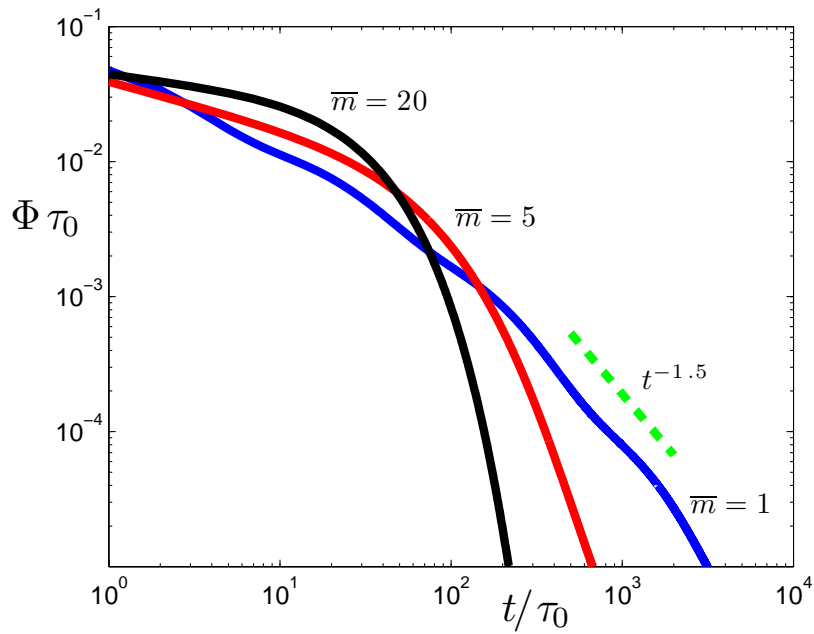}
\caption{(Color online).  Departure time
distribution function versus time as determined by Eq. \protect\ref
{deptimedistdif} in the diffusive regime. \ In the plot the average binding energy 
is held constant at $3k_{B}T$. The theoretically determined departure time distribution can be
compared to an experiment with DNA-grafted colloids (see Section VI) which
observed power law behavior with exponent $-1.5$. \  }
\label{coarsedif2}
\end{figure}

In Fig. \ref{coarsedif2} the departure time distribution is plotted in the
diffusive regime. \ The optimal regime for fast particle departure is to
have a large number ($\overline{m}\sim 10$) of weakly bound key-lock
bridges. \ In this scenario the departure time distribution is accurately
approximated as a single exponential, $\Phi (t)=K_{\overline{m}}\exp (-K_{%
\overline{m}}t)$. \ 

\section{Diffusion}

We now turn to discuss the statistics for the in-plane diffusion of the
particle. \ We first note that the in-plane trajectory of the particle
subjected to a delta-correlated random potential remains statistically
equivalent to an unbiased random walk. \ As a result, the mean squared
displacement for an $n$ step random walk remains $\left\langle
r^{2}\right\rangle =na^{2}$. \ As the particle explores the landscape it
cascades into deeper energy wells, the hopping time increases, and the
diffusion gets slower. \ In the limit $n\rightarrow\infty$ the average
hopping time can be determined from the equilibrium canonical distribution.
\ For Poisson distributed bridge numbers $m$, this corresponds to a finite
renormalization of the diffusion coefficient $D^{\ast}$ with $%
D_{0}=a^{2}/4\tau_{0}$. \ 

\begin{figure}[tbp]
\includegraphics[width=4.6138in,height=3.4714in]{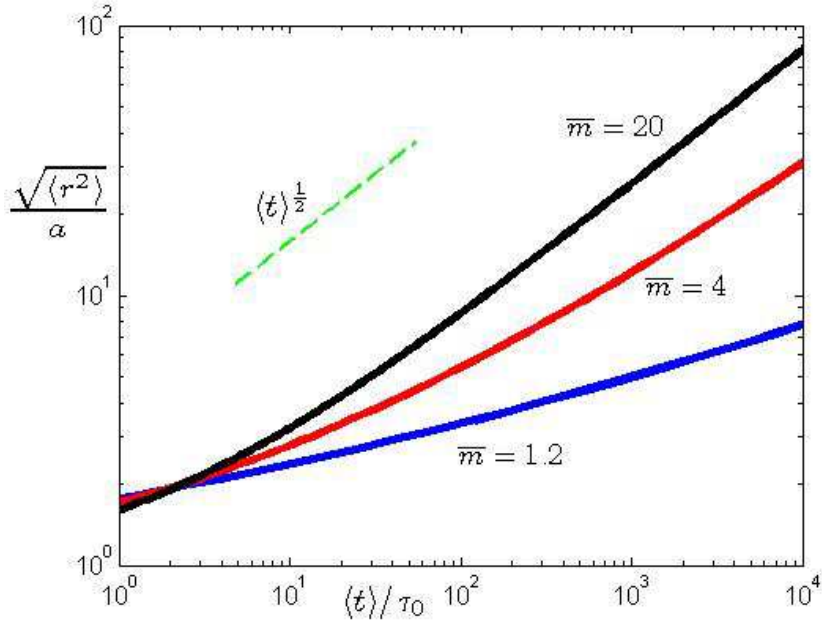}
\caption{(Color online). \ Root mean squared displacement vs. time with $%
\Delta\overline{m}=4$. \ The curves are calculated from the parametric
equations \protect\ref{rsquared}, \protect\ref{tav}. \ }
\label{rvtnew}
\end{figure}

\begin{figure}[tbp]
\includegraphics[width=4.6138in,height=3.4714in]{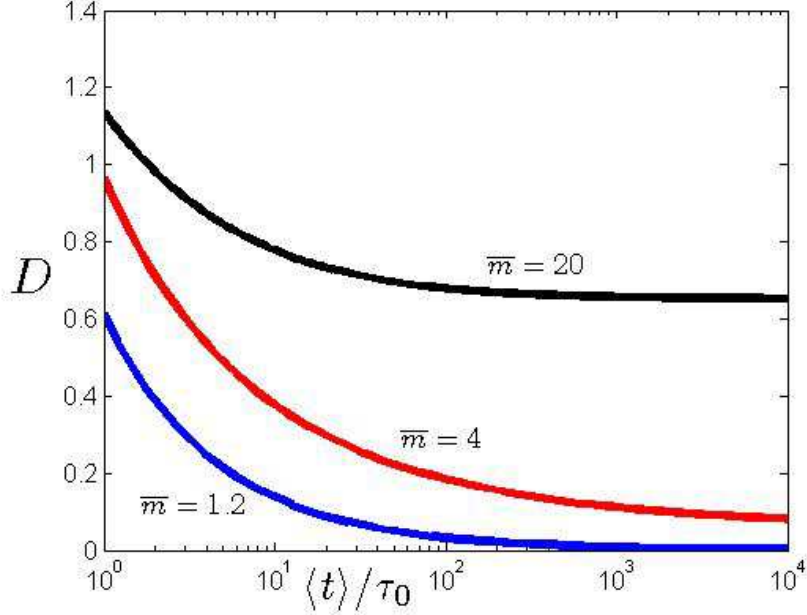}
\caption{(Color online). \ The dimensionless diffusion coefficient $D\equiv%
\frac{1}{4D_{0}}\frac{\partial\left\langle r^{2}\right\rangle }{%
\partial\left\langle t\right\rangle }$ plotted against time. \ }
\label{Dnew}
\end{figure}

\ 
\begin{equation}
\left\langle t\right\rangle =n\left\langle \tau_{m}\right\rangle =n\tau
_{0}\exp\left( -\Delta\overline{m}\right) \sum_{m=1}^{\infty}\widetilde {P}%
_{m}\exp\left( \Delta m\right) =n\tau_{0}\frac{\exp\left( \overline {m}%
e^{\Delta}\right) -1}{\exp\left( \Delta\overline{m}\right) [\exp(\overline{m}%
)-1]}
\end{equation}%
\begin{equation}
D^{\ast}\equiv\frac{1}{4}\frac{\partial\left\langle r^{2}\right\rangle }{%
\partial\left\langle t\right\rangle }=D_{0}\frac{\exp\left( \Delta \overline{%
m}\right) [\exp(\overline{m})-1]}{\exp\left( \overline{m}e^{\Delta}\right) -1%
}
\end{equation}

This "ergodic" behavior is only achieved after a very long time. \
Generally, an $n$ step random walk cannot visit sites with arbitrarily large 
$m$. \ In this transient regime one should only average over sites with $%
m<m^{\ast}$. \ In the language of the statistics of extreme events, $%
m^{\ast}-1$ is the maximum "expected" value of $m$ in a sample of $n$
independent realizations\cite{dispersive}. \ Even with this complication,
the average diffusion time $\left\langle t\right\rangle $ and the mean
squared displacement $\left\langle r^{2}\right\rangle $ can both be
expressed in terms of $m^{\ast}$, which defines their relationship in
parametric form. \ \ 
\begin{equation}
\left\langle r^{2}\right\rangle =\frac{a^{2}}{P(\overline{m},m^{\ast})}
\label{rsquared}
\end{equation}%
\begin{equation}
\left\langle t\right\rangle =\frac{\left\langle r^{2}\right\rangle }{D^{\ast}%
}\left( 1-\frac{P(\overline{m}e^{\Delta},m^{\ast})}{1-\exp(-\overline {m}%
e^{\Delta})}\right)  \label{tav}
\end{equation}

Here $P(x,m^{\ast })\equiv \gamma (x,m^{\ast })/\Gamma (m^{\ast })=\exp
(-x)\sum_{k=m^{\ast }}^{\infty }x^{k}/k!$ is the regularized lower
incomplete $\Gamma $ function. \ In the limit $m^{\ast }\rightarrow \infty $
we recover the renormalized diffusion relation $\left\langle t\right\rangle
=\left\langle r^{2}\right\rangle /D^{\ast }$, although this occurs at very
long, often unrealistic times. \ In the transient regime we expect
anomalous, subdiffusive behavior. \ As indicated in Fig. \ref{rvtnew}, this
subdiffusive behavior is typical for strong enough key-lock interactions. \
Figure \ref{Dnew} is a plot of the dimensionless diffusion coefficient
versus time. \ The anomalous diffusion may be well described by a power law $%
\left\langle r^{2}(t)\right\rangle \sim \left\langle t\right\rangle ^{\alpha
}$ with a nonuniversal exponent $\alpha <1$. \ 

There is an analogy between our results and dispersive transport in
amorphous materials. \ In these systems a length dependence of the effective
mobility\cite{semiconductortransport} can be interpreted within the context
of the statistics of extreme events\cite{dispersive}. \ The time required
for charge carriers to travel through the material depends on the dwell
times spent at all of the trapping centers. \ Since this transit time will
be dominated by the dwell time of the deepest trapping center, one would
like to know how the thickness of the material effects the distribution for
the largest trapping depth. \ The analogy to our result is made by replacing
the material thickness with the number of steps in the random walk $n$, and
replacing the distribution for the largest trap depth with the distribution
for $m^{\ast }$, since the bridge number is related to energy by Eq. \ref%
{potenergy}. \ 

\section{Experiment}

We now would like to make a connection between our results and recent
experiments with DNA-grafted colloids. \ The departure time distribution can
be compared to an experiment which determined the time-varying separation of
two DNA-grafted colloids in an optical trap\cite{crocker}. \ In the
experimental setup, two particles are bound by DNA bridges, and after
breaking all connections diffuse to the width of the optical trap. \ Because
the length of the DNA\ chains grafted on the particle is much shorter than
the particle radius, surface curvature effects can be neglected. \ The
interaction resembles that of a particle interacting with a patch on a $2D$
substrate. \ Experimentally the tail of the departure time distribution was
observed to be a power law $\Phi (t)\sim t^{-1.5}$. \ Qualitatively similar
behavior is predicted by the theory with $\overline{m}\sim 1$ and average
binding free energy of several $k_{B}T$ (see $\overline{m}=1$ curve in Fig. \ref{coarsedif2}). \ 

\begin{figure}[tbp]
\includegraphics[width=4.6112in,height=3.4705in]{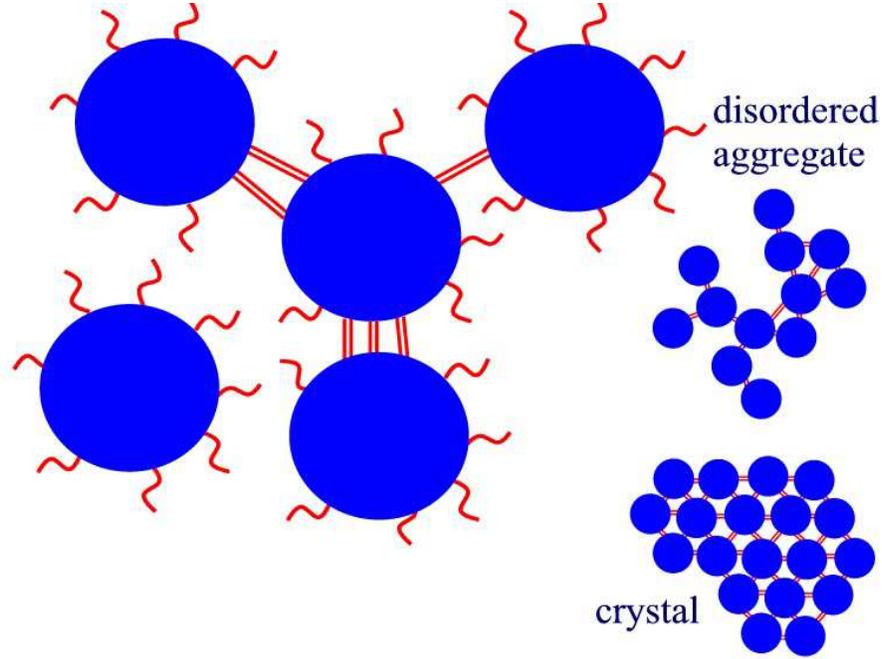}
\caption{(Color online). \ Graphical depiction of key-lock binding between
nanoparticles functionalized with complementary ssDNA. \ The resulting
structures can be disordered, fractal-like aggregates, or crystalline. \ }
\label{crystal}
\end{figure}

In addition, our work provides insight into the slow crystallization
dynamics of key-lock binding particles (see Fig. \ref{crystal}). \ In \cite%
{chaikin}, $1\mu m$ diameter particles grafted with ssDNA formed reversible,
disordered aggregates. \ The average number of key-lock bridges between
particles was $\overline{m}\sim 2$. \ The authors of \cite{crocker} observed
random hexagonal close packed crystals by further reducing the surface
density of DNA strands on the particles. \ The crystallization process
requires that particles rapidly detach and reattach at the desired lattice
location. \ In the localized regime particle desorption is the relevant
process. \ 

\begin{figure}[tbp]
\includegraphics[width=4.6138in,height=3.4714in]{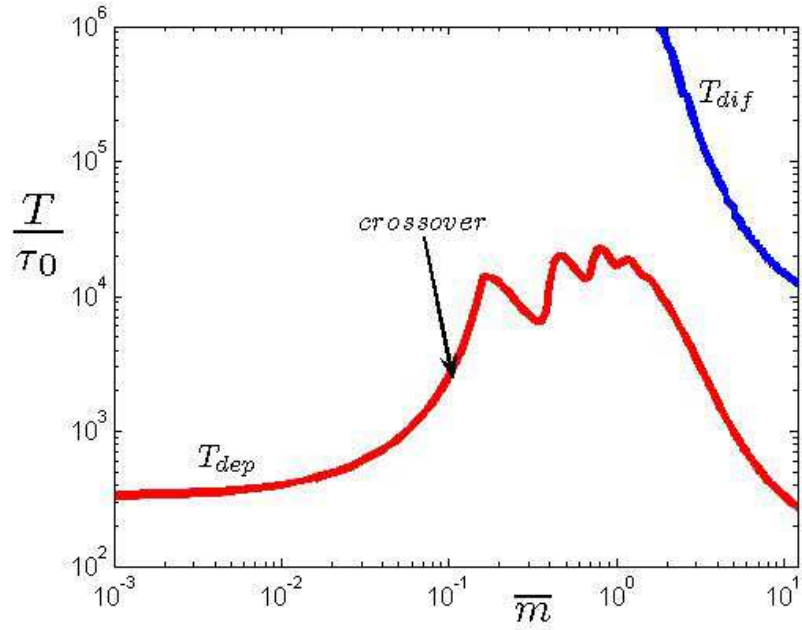}
\caption{(Color online). \ Plot of the
characteristic times $T_{dep}$ and $T_{dif}$ versus $\overline{m}$ at
constant binding energy ($const=4$ in Eq. \protect\ref{constbinding}). \  }
\label{timecompare}
\end{figure}

In the diffusive regime surface diffusion also plays a role in the
rearrangement of particles into the desired crystalline structure. \ To
determine which process is more important for particle rearrangement, we can
compare the departure time with the time required for a particle to
diffusively explore the surface of a particle to which it is bound. \ The
time $T_{dep}$ required for $90\%$ of the particles to depart is: \ \ 
\begin{equation}
0.1=\int_{T_{dep}}^{\infty }\Phi (t)dt  \label{Tdep}
\end{equation}

To estimate the time required for diffusive rearrangement $T_{dif}$ we use
the parametric equations \ref{rsquared},\ref{tav}. \ In \cite{chaikin}
particles of radius $R=.5\mu m$ were grafted with DNA chains of length $%
l\sim 20nm$. \ Assuming the correlation length $a\sim l$ we have $\frac{%
\left\langle r^{2}\right\rangle }{a^{2}}\sim \left( \frac{\pi R}{l}\right)
^{2}\simeq 10^{3}$. \ Figure \ref{timecompare} shows a comparison of $%
T_{dif} $ and $T_{dep}$ at constant binding free energy. \ 
\begin{equation}
\frac{\Delta \overline{m}}{1-\exp (-\overline{m})}+\log (1-\exp (-\overline{m%
}))=const  \label{constbinding}
\end{equation}

This expression for the binding energy takes into account the entropy
reduction associated with the non-ergodic degrees of freedom. \ For a
detailed discussion of this topic see reference \cite{statmech}. \ Since $%
T_{dif}>T_{dep}$, colloidal desorption and reattachment is the dominant
mechanism by which particles rearrange. \ 

\ As the figure indicates, the optimal regime of fast departure is to have a
large number ($\overline{m}\gtrsim 10$) of weakly bound key-lock bridges. \
We predict a localized regime below the crossover where particle departure
is relatively fast. \ Just beyond the crossover there is a relative maximum
in $T_{dep}$ before it decreases at large $\overline{m}$. \ The increase in
departure time at the onset of diffusive behavior is indicative of a regime
where the system ages. \ In this regime the interplay of diffusion an
desorption leads to longer bound state lifetimes, and an increase in the
departure time. \ 

\begin{figure}[tbp]
\includegraphics[width=4.6112in,height=3.4705in]{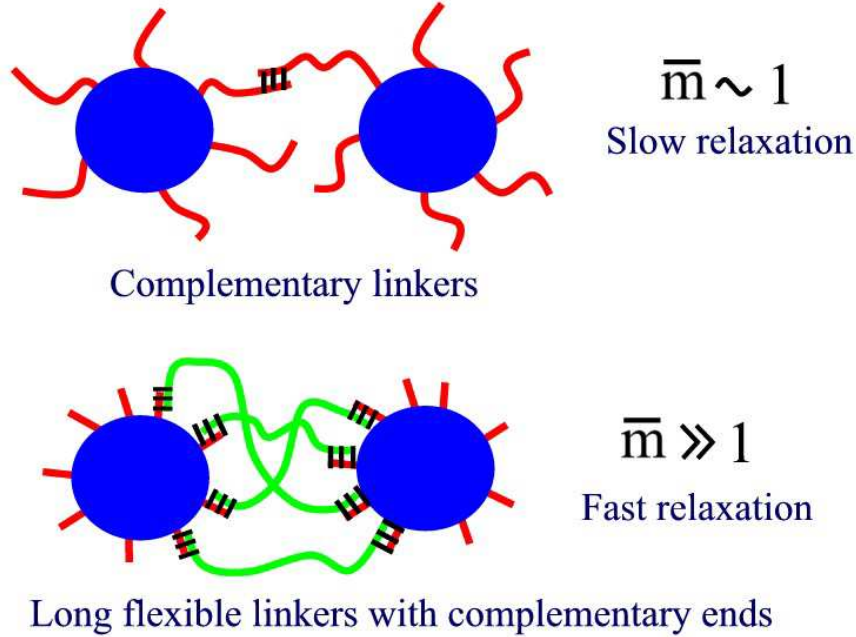}
\caption{(Color online). \ Graphical depiction comparing recent experiments
with DNA-grafted colloids to a future implementation with long flexible
linkers. \ Increasing the number of key-lock bridges between particle pairs
potentially decreases the time required for crystallization. \ }
\label{longlinkers}
\end{figure}

We now turn to the question of designing a future experiment which will
facilitate fast particle departure and hence colloidal crystallization. \
The essential goal is to increase the average number of key-lock bridges
between particle pairs. \ Increasing the surface density of DNA strands
alone results in the formation of a brush, which decreases the effective
cross section for the interaction between complementary DNA. \ Instead we
propose the introduction of long, flexible DNA\ linkers between particles
with a high coverage of short ssDNA (see Fig. \ref{longlinkers}). \ This
system has the potential to realize more key-lock bridges between particle
pairs as compared to previous experiments, and therefore substantially
reduce the time required for crystallization. \ 

\section{Conclusion}

This paper studied the dynamics of particles which interact through the
reversible formation of multiple key-lock bridges. \ Well before the
percolation threshold is reached there is a crossover from a localized
regime to a diffusive regime. \ In the localized regime the particles remain
close to their original attachment site until departing. \ In this regime
particles are attached to finite clusters, and the system exhibits an
exponential distribution of departure times. \ Once the radius of gyration
of the cluster exceeds the characteristic radius for the particles' random
walk, the finite clusters behave effectively as infinite clusters. \
Diffusion allows the particles to cascade into deeper energy wells, which
leads to a decrease in hopping rate. \ The diffusion slows and the bound
state lifetime increases, a phenomenon qualitatively similar to \textit{aging%
} in glassy systems. \ In the diffusive regime we discussed the statistics
for the particles' in-plane diffusion. \ Weak key-lock interactions give
rise to a finite renormalization of the diffusion coefficient. \ However, as
the strength of the interaction increases (larger $\Delta $), the system
exhibits anomalous, subdiffusive behavior. \ This situation is analogous to
dispersive transport in disordered semiconductors. \ We then made the
connection between our calculation of the departure time distribution and
recent experiments with DNA-coated colloids. \ The findings indicate that
the optimal regime for colloidal crystallization is to have a large number
of weakly bound key-lock bridges. \ A modified experimental setup was
proposed which has the potential to realize this regime of fast particle
departure. \ 

\begin{acknowledgments}
This work was supported by the ACS Petroleum Research Fund (PRF Grant No.
44181-AC10). \ We acknowledge L. Sander, B. Orr, and B. Shklovskii for
valuable discussions. \ 
\end{acknowledgments}

\bibliographystyle{apsrev}
\bibliography{acompat,dna}

\end{document}